\documentclass[pre,amssymb,amsmath,floats,twocolumn,showpacs,superscriptaddress]{revtex4}

\usepackage{epsfig}

\begin{document}

\title{Phase Transition with the Berezinskii--Kosterlitz--Thouless Singularity 
\\
in the Ising Model on a Growing Network}

\author{M. Bauer}
\email{michel.bauer@cea.fr}
\affiliation{Service de Physique Th\'eorique, 
%%de Saclay CEA/DSM/SPhT, Unit\'e de recherche associ\'ee au 
CNRS CEA-Saclay, 91191 Gif-sur-Yvette, France}

\author{S. Coulomb}
\email{coulomb@dsm-mail.saclay.cea.fr}
\affiliation{Service de Physique Th\'eorique, 
%%de Saclay CEA/DSM/SPhT, Unit\'e de recherche associ\'ee au 
CNRS CEA-Saclay, 91191 Gif-sur-Yvette, France}

\author{S. N. Dorogovtsev}
\email{sdorogov@fis.ua.pt}
\affiliation{Departamento de F{\'\i}sica da Universidade de Aveiro, 3810-193 Aveiro, Portugal}
\affiliation{A. F. Ioffe Physico-Technical Institute, 194021
  St. Petersburg, Russia}

%%\date{\today}
\date{}

\begin{abstract} 

We consider the ferromagnetic Ising model on a highly inhomogeneous network 
created by a growth process. We find that the phase transition in this system is characterised by the Berezinskii--Kosterlitz--Thouless singularity, although critical fluctuations are absent and the mean-field description 
%%in the framework of a mean-field approach 
is exact.  
Below this infinite order 
%%phase 
transition, the magnetization behaves as $\exp(-\text{const}/\sqrt{T_c-T})$. 
%%All the derivatives of the specific heat approach zero at the critical point, where the susceptibility has a finite jump. 
We show that the critical point separates the phase with the power-law distribution of the linear response to a local field
%%%---an analogy of a pair correlation function for lattices---
and the phase where this distribution rapidly decreases. 
We suggest that this phase transition occurs in a wide range of cooperative models with a strong 
%%long
infinite-range 
inhomogeneity. 
%%%%%%%%%%%%%%%%%%%%%%%
%%%%%%%%%%%%%%%%%%%%%%%
%%%%%%%%%%%%%%%%%%%%%%% 
\\
{\em Note added.}---After this paper had been published, we have learnt that the infinite order phase transition in the effective model we arrived at was discovered by O.~Costin, R.D.~Costin and C.P.~Gr\"unfeld in 1990. This phase transition was considered in the following papers:  
\\
$[$1$]$    
O.~Costin, R.D.~Costin and C.P.~Gr\"unfeld, 
Infinite-order phase transition in a classical spin system,  
J. Stat. Phys. {\bf 59}, 1531 (1990);
\\
$[$2$]$  
O.~Costin and R.D.~Costin, 
Limit probability distributions for an infinite-order phase transition model,  
J. Stat. Phys. {\bf 64}, 193 (1991);   
\\
$[$3$]$ 
M.~Bundaru and C.P.~Gr\"{u}nfeld, 
On a phase transition in a one-dimensional non-homogeneous model, 
J. Phys. A {\bf 32}, 875 (1999);  
\\
$[$4$]$ 
S.~Romano, 
Computer simulation study of one-dimensional lattice spin models with long-range inhomogeneous interactions, 
Mod. Phys. Lett. B {\bf 9}, 1447 (1995). 
\\
We would like to note that Costin, Costin and Gr\"unfeld treated this model as a one-dimensional inhomogeneous system. We have arrived at the same model as a one-replica ansatz for a random growing network where expected to find a phase transition of this sort based on earlier results for random networks (see the text). We have also obtained the distribution of the linear response to a local field, which characterises correlations in this system.    
We thank O. Costin and S. Romano for indicating these publications of 90s. 
%%%%%%%%%%%%%%%%%%%%%%%
%%%%%%%%%%%%%%%%%%%%%%%
%%%%%%%%%%%%%%%%%%%%%%%   
\end{abstract}


\maketitle

The ferromagnetic Ising model on lattices has an ordinary second-order phase transition \cite{o44}. Above the upper critical dimension of the model, 
%%(four), 
the critical fluctuations are absent, and the mean-field description of this transition is exact. In particular, this takes place if couplings between all spins are equal---infinite-range interactions. 
%%%%%%%%%!!!!!These substrates (fully connected graphs or ``infinitely-dimensional lattices'') were mostly used for studying more complex spin-glass models \cite{sk75}. 
%%%%%%%%%!!!!!
In this Letter we report the finding of a phase transition with the Berezinskii--Kosterlitz--Thouless (BKT) singularity in the Ising model 
on a growing network, which is infinitely-dimensional as most of networks.  
This transition is quite unusual for an infinitely-dimensional system  
%%, and networks, generally, are infinitely-dimensional. 
%%(As a rule, networks are infinitely-dimensional objects in the sense that the mean intervertex distances of networks grow slower than any positive power of the numbers of their vertices.) 
as well for a cooperative model with the order parameter of discrete symmetry. 

Recall that in ``ordinary'' continuous phase transitions, 
pair correlations of an order parameter show a slow, power-law space decay only at the critical point, $T_c$, and decay exponentially both in the low- and high-temperature phases. This behavior was observed in the Ising model on equilibrium complex networks \cite{ahs02,dgm02,lvv02,g03,remark1} (for percolation and for disease spreading on equilibrium networks, see Refs.~\cite{cah02} and \cite{pv01}, respectively).  
%%%%(The complex architectures of the equilibrium networks may change the values of critical exponents and increase the order of the phase transition. 
%%%%However, they do not change dramatically the type of critical singularity.) 
%%, but these are not so impressive changes as the emergence of a new kind of a critical singularity.) 
%%%Although the complex architectures of these networks change .... 
%%
In contrast, the BKT phase transition \cite{b71,kt73} separates the phase with rapidly decreasing correlations and the critical phase 
%%(``the line of critical points'', as it is often called) 
with correlations decaying by 
a power law. 
The contact of these two phases is characterised by specific 
%%critical 
dependences. For example, the order parameter behaves as $M(T) \sim \exp(-\text{const}/\sqrt{T_c-T})$, and 
%%so 
the phase transition is of infinite order.  

%%Without going into detail, t
Normally, the BKT transition is realized in systems with two-component order parameters of continuous symmetry at a lower critical dimension. 
Also, this ano\-ma\-lous phase transition is present in a few low-dimensional systems (e.g., the Luttinger liquid) which actually can be reduced to above indicated ones. 
There is one more interesting situation, where the BKT singularity emerges. 
It was observed that in some growing networks, near the birth point of the giant connected component, its relative size behaves similarly to the magnetization near the BKT transition    
%%A phase transition of the birth of the giant connected component with the BKT singularity was earlier observed in specific growing networks 
\cite{chk01,dms01,kkkr02,l02,bb03,d03,cb03,br05,kd04}. 

%%In the present Letter, w
{\em The model.}---We find an exact solution of the following cooperative model. 
%%We grow a network up to a large size and 
A network grows up to a large size, and 
%%then spins are placed on vertices of the network. 
and interacting spins are considered on the resulting net.  
The interaction between the nearest neighbor spins is described by the ferromagnetic Ising model. 

%%%%%%%%%%%%%%%%%%%%%%%%%%%%%%%%%%%%%%%%%%%%%%%%%%%%%%%%%%%%%%%%%%%%
%%%%%%%%%%%%%%%%%%%%%%%%%%%%%%%%%%%%%%%%%%%%%%%%%%%%%%%%%%%%%%%%%%%%

\begin{figure}[b]
\epsfxsize=60mm
%%72mm %58mm
\epsffile{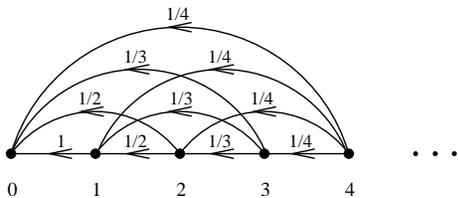}
\caption{ 
Deterministic system of interacting spins which is equivalent to our growing network. 
%%model. 
The numbers on the edges show the values of the Ising couplings between the spins at the corresponding vertices. 
}
\label{f1}
\end{figure}

%%%%%%%%%%%%%%%%%%%%%%%%%%%%%%%%%%%%%%%%%%%%%%%%%%%%%%%%%%%%%%%%%%%%
%%%%%%%%%%%%%%%%%%%%%%%%%%%%%%%%%%%%%%%%%%%%%%%%%%%%%%%%%%%%%%%%%%%% 

%%Our growing network is very simple:  
We use the following growing \vspace{-10pt}network: 

\begin{itemize}

\item[(i)] 
The growth starts with a single vertex \vspace{-9pt}($t=0$). 

\item[(ii)]  
At each time step, we add a new vertex and attach it to one of ``older'' \vspace{-9pt}vertices. 
%%chosen with equal probability. 

\item[(iii)]  
For simplicity, we suggest a specific annealing. For an edge born at time $t$, the end of the edge at vertex $t$ is fixed, and the second end can be found at each of the vertices in the range $0\leq\tau<t$ with equal probability. Characteristic times for the jumps of this end between the vertices $0\leq\tau<t$ are assumed to be not greater than those of the spin \vspace{-10pt}relaxation.

\end{itemize}

One can show that the resulting model is equivalent to the ferromagnetic Ising model on the deterministic graph shown in Fig.~\ref{f1}. 
In this system, the spin on a vertex, which was born at time $t$, has equal coupling $1/t$ to each of spins on the older vertices. 
The Hamiltonian of the model is: 
\begin{equation} 
{\cal H} =\, -\!\!\!\!\sum_{0\leq i<j\leq t} \frac{s_i s_j}{j} - \sum_{i=0}^t H_i s_i
\, ,  
\label{e1}
\end{equation} 
where spins $s_i=\pm 1$, and $H_i \geq 0$ is an inhomogeneous magnetic field. 
Actually, we reduce our problem to a system with a strong deterministic infinite-range disorder. 
%%, where all vertices interact with each other with different strength.  

{\em Mean-field treatment.}---Let us first use a mean-field ansatz. 
We will show afterwards that the mean-field solution is exact. For the sake of brevity, we use the following simple mean-field treatment. We assume small fluctuations of spins from their mean-field values $m_i$: 
$s_i s_j \to m_i m_j + m_i(s_j-m_j) + m_j(s_i-m_i)$. Substituting this relation into Eq.~(\ref{e1}) gives a linear effective mean-field Hamiltonian. With this Hamiltonian, it is easy to obtain the partition function $Z = \sum_{\{s_i=\pm1\}}e^{-\beta{\cal H}[\{s_i\}]}$ ($\beta \equiv 1/T$) and the free energy $F = - \beta^{-1}\ln Z$:
\begin{eqnarray} 
&&
F = t\! \int_0^1 \!dx \int_x^1 \frac{dy}{y}\,m(x)m(y) - \frac{t}{\beta}\ln2- 
\nonumber
\\[5pt]
&&
\phantom{|}\!\!\!\!\!\!\!\!\!\!\!
\frac{t}{\beta} \!\!\int_0^1 \!\!\!dx\ln \cosh\! 
\left\{\!\beta\!\left[\frac{1}{x}\int_0^x\!\!\!\!\! dy\,m(y) + \!\int_x^1 \!\!\frac{dy}{y}\,m(y) + H(x)\right]\right\} .
\nonumber
\\[5pt]
&&
%%\,   
\label{e2}
\end{eqnarray} 
Here we assumed that $t$ is large and passed to the continuum limit: 
$m_i = m(x\!=\!i/t)$. Expression (\ref{e2}) together with the relation 
$m(x) = -(1/t)\,\delta F/\delta H (x)$ 
allows us to obtain the equation for the mean local magnetization: 
\begin{equation} 
m(x) = \tanh
\left\{\!\beta\!\left[\frac{1}{x}\int_0^x\!\!\!\! dy\,m(y) + \!\int_x^1 \!\frac{dy}{y}\,m(y) + H(x)\right]\right\}
\, .   
\label{e3}
\end{equation} 

{\em The exact derivation of the free energy.}---The free energy can be found exactly. We compare the free energies of the network at times $t-1$ and $t$. For brevity, here we consider only the homogeneous magnetic field. 
The form of the Hamiltonian (\ref{e1}) results in the relation  
%%Using form (\ref{e1}) of the Hamiltonian of our model gives 
%%
\begin{equation} 
e^{-\beta F_t(H)} = \sum_{s_{t}=\pm 1} e^{-\beta F_{t-1}(H+s_{t}/t)}e^{\beta Hs_{t}} 
\, , 
\label{e4}
\end{equation} 
that is,  
\begin{eqnarray} 
&&
e^{-\beta [F_t(H)-F_{t-1}(H)]} \to e^{-\beta F_t(H)/t} = 
\nonumber
\\[5pt]
&&
\sum_{s=\pm 1} \exp\left[-\beta \frac{\partial F_{t}(H)}{\partial H}\,\frac{s}{t} + \beta Hs\right]
\, . 
\label{e5}
\end{eqnarray} 
Here we took into account the fact that at large $t$, the ratio $F_t(H)/t$ approaches a $t$-independent limit. Using the relation for the (relative) 
full magnetization   
$M(H) = \int_0^1 dx\, m(x,H) = - (1/t)\partial F_{t}(H)/\partial H$, we get the exact form of the free energy: 
\begin{equation} 
F = - t \,\beta^{-1} \ln\{ 2\cosh[\beta(H+M(H))]\}
\, 
%%. 
\label{e6}
\end{equation} 
at $t \to \infty$. 
One can check that free energy expressions (\ref{e2})---the mean-field one---and (\ref{e6})---the exact expression---coincide.  
Indeed, substituting Eq.~(\ref{e3}) into the relation (\ref{e2}) and making partial integration, we arrive at the free energy exactly in form~(\ref{e6}). 
In this sense, the mean-field treatment of this problem is exact. 
%%This claim is also fulfilled in the case of inhomogeneous magnetic field.  

{\em Analysis of the equation for the magnetization.}---Let us consider Eq.~(\ref{e3}). From this equation, one can 
%%immediately 
see that the assumption $m(x) \neq 0$ at some 
%%value of 
$x$ immediately leads to the following behavior of $m(x)$ near $x=0$: 
\begin{equation} 
m(x \sim 0) \cong 1 - A\,x^{2\beta}
\, ,    
\label{e7}
\end{equation} 
where $A$ depends on $\beta$ and $H$.  
If $H=0$, this behavior is realized only in the low-temperature phase, and $m(x)=0$ above $T_c$. If $H> 0$, $m(x=0)=1$ at any temperature. 
%%(We suggest that the critical point, $T_c$, exists. 
(We will see that the critical point, $T_c$, exists.) 
%%this is true.) 
%The resulting magnetization profile $m(x)$ is schematically shown in Fig.~\ref{f2}. 
%%($T<T_c$ and $H=0$). 
On the boundary, Eq.~(\ref{e3}) readily gives $m(1) = \tanh\{\beta [M+H(1)]\}$. 
%%From Eq.~(\ref{e3}), one can see that on the boundary, $m(1) = \tanh[\beta M+H(1)]$. 

%%%%%%%%%%%%%%%%%%%%%%%%%%%%%%%%%%%%%%%%%%%%%%%%%%%%%%%%%%%%%%%%%%%%
%%%%%%%%%%%%%%%%%%%%%%%%%%%%%%%%%%%%%%%%%%%%%%%%%%%%%%%%%%%%%%%%%%%%

%\begin{figure}[b]
%\epsfxsize=52mm
%%\begin{center}
%\epsffile{bkt_figure2.eps}
%%\end{center}
%\caption{ 
%Schematic plot of the magnetization profile $m(x)$ in the low-temperature phase, $\beta\equiv1/T>\beta_c=1/4$. 
%$x=i/t$ is a relative birth-time of a vertex. Magnetic field is absent. 
%As temperature approaches the critical one, the profile decreases with $x$ more and more rapidly, and at the phase transition point, $m(x)$ approaches zero. 
%}
%\label{f2}
%\end{figure}

%%%%%%%%%%%%%%%%%%%%%%%%%%%%%%%%%%%%%%%%%%%%%%%%%%%%%%%%%%%%%%%%%%%%
%%%%%%%%%%%%%%%%%%%%%%%%%%%%%%%%%%%%%%%%%%%%%%%%%%%%%%%%%%%%%%%%%%%% 

For finding this profile, it is convenient to pass to a differential equation. 
For brevity, here we assume that $H=0$. 
We introduce a new variable, $n(z)$: 
\begin{equation} 
%%\widetilde{m}(1/x) \equiv \frac{1}{x}\int_0^x\!\! dy\,m(y) + \int_x^1 \frac{dy}{y}\,m(y)
n(-\ln x)\equiv\beta\left[ \frac{1}{x}\int_0^x\!\! dy\,m(y) + \int_x^1 \frac{dy}{y}\,m(y)\right]
\, ,    
\label{e8}
\end{equation} 
so 
%%that 
%%
\begin{equation} 
%%m(x) = \tanh[\beta\widetilde{m}(1/x)]
m(x) = \tanh n(-\ln x)
\, .    
\label{e9}
\end{equation} 
%% 
%%Differentiating $\widetilde{m}(x)$ two times 
Differentiating Eq.~(\ref{e8}) and using Eq.~(\ref{e9}) gives the second order differential equation 
\begin{equation} 
%%\frac{d^2\widetilde{m}(x)}{d x^2} = -\frac{1}{\phantom{.}x^2}\,\tanh[\beta\,\widetilde{m}(x)]
\frac{dn(z)}{dz} - \frac{d^2n(z)}{dz^2} = \beta\tanh n(z)
\, 
%%.    
\label{e10}
\end{equation} 
with the 
%%following 
boundary conditions:    
%%($1 \leq x < \infty$): 
%%$(d\widetilde{m}/dx)(x\!=\!1) = \widetilde{m}(x\!=\!1) = M \equiv \int_0^1 dy\,m(y)$ 
%%and 
%%$\widetilde{m}(x\to\infty) \cong \text{const} + \ln\,x$. 
%%The boundary condition at $x=1$ follows from definition (\ref{e8}). 
%%The boundary condition at $x\to\infty)$ follows from relation (\ref{e7}). 
%%($z$ is related to $x=i/t$ in the following way: $z=-\ln x$, so $0 \leq z < \infty$, where $z=0$ corresponds to $x=1$): 
(i) $(dn/dz)(z\!=\!0) = n(z\!=\!0)$ [note that $n(z\!=\!0) = \beta M \equiv \beta\int_0^1 dy\,m(y)$] 
and 
(ii) $n(z\to\infty) \cong \beta z + \text{const}$. 
$z$ is related to $x=i/t$: 
%%in the following way: 
$z=-\ln x$, so $0 \leq z < \infty$, where $z=0$ corresponds to $x=1$. 
Boundary conditions (i) and (ii) follow from definition (\ref{e8}) and  
%%Boundary condition (ii) follows from 
relation (\ref{e7}), respectively.  
%%Note that a
At each value of $\beta$, there is a single solution of Eq.~(\ref{e10}) with these boundary conditions, which allows one to get $M$.  

Equation (\ref{e10}) can be transformed into a first order differential equation. For this, we pass from variables $\{t,n(t)\}$ to $\{n,w(n)\}$, where 
$w \equiv \beta^{-1}(dn/dt)$. 
[$n$ varies from $0$ to $\infty$, while $w(n)$ takes values between $0$ and $1$.] 
This gives the equation 
\begin{equation} 
w\frac{dw}{dn} = \beta^{-1}(w - \tanh n)
\, 
%%.    
\label{e11}
\end{equation} 
for $w(n)$ with the following boundary conditions: 
(i) $w[n(z\!=\!0)]=\beta^{-1}n(z\!=\!0)$ [recall that $\beta^{-1}n(z\!=\!0) = M$] and (ii) $w(n\to\infty)=1$. 
Here, boundary condition (i) on the line $w=\beta^{-1}n$ 
corresponds to that at %%a newborn vertex. 
$x=i/t=1$.
Asymptotic boundary condition (ii) corresponds to the limit $i/t \to 0$. 
Knowing 
%%the solution 
$w(n)$ one can easily get $m(x)$. 

The analysis of Eq.~(\ref{e11}) is similar to that of an equation of this type in Ref.~\cite{dms01}. 
At small $n$, one can substitute $\tanh n$ by $n$ on the right-hand side of Eq.~(\ref{e11}), so we have 
$w dw/dn = \beta^{-1}(w - n)$.  
%%%%
%%\begin{equation} 
%%w\frac{dw}{dn} = \beta^{-1}(w - n)
%%\, .     
%%\label{e11a}
%%\end{equation} 
%%%% 
The solutions of 
this
%%the resulting 
equation can be presented in an analytical form. A physically reasonable non-zero solution must cross the ordinate axis (and the $w=\beta^{-1}n$ line) 
at non-negative $w$. This solution exists if $\beta\geq1/4$. 
%%, which indicates the presence of a phase transition. 
%%At the 
There is a critical point, $\beta_c=1/4$, where the solution is 
\begin{equation} 
w_c(n,\beta=1/4) = 2n[1-f(n)] 
\, 
%%.      
\label{e11a}
\end{equation}
with $f(n)$ satisfying $f(n\to 0)\to 0$ and the 
%%following 
relation: 
$\ln[nf(n)] + 1/f(n) = \ln c$. 
Here the constant $c=1.554\ldots$ ensures that $w_c(n)$ (\ref{e11a}) fits the corresponding solution of Eq.~(\ref{e11}) 
which approaches $1$ as $n \to \infty$. 
The form of the critical solution at small $n$ indicates the presence of the BKT singularity. 

\begin{figure}[t]
%%\epsfxsize=61mm
%%\epsffile{bkt_figure3a.eps}
%%\epsfxsize=61mm
%%\epsffile{bkt_figure3b.eps}
%%\epsfxsize=61mm
%%\epsffile{bkt_figure3c.eps}
\begin{center}
\scalebox{0.261}{\includegraphics[trim=0bp 0bp 0bp -35bp,angle=270]{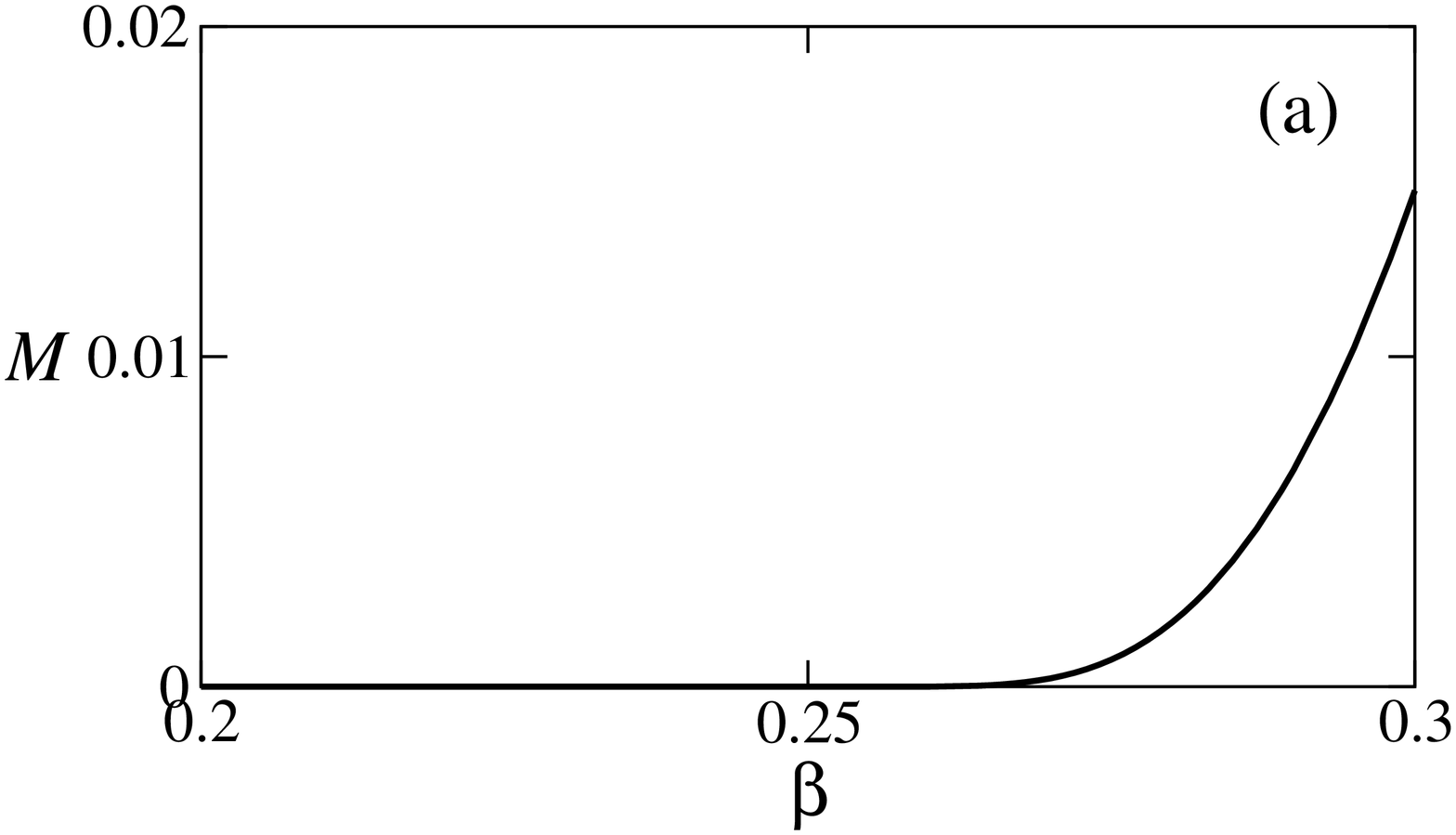}}
\scalebox{0.24}{\includegraphics[angle=270]{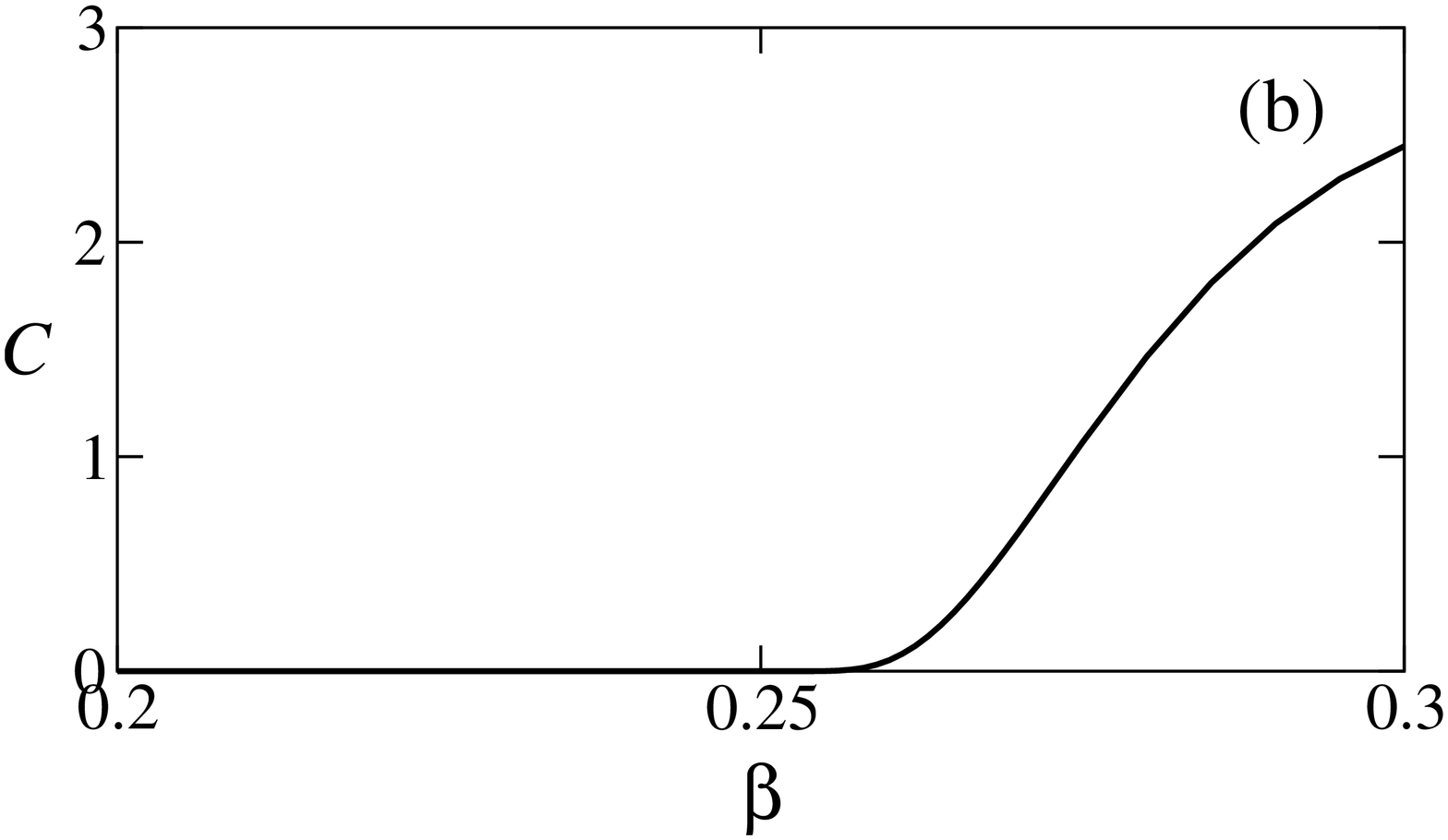}}
\scalebox{0.24}{\includegraphics[angle=270]{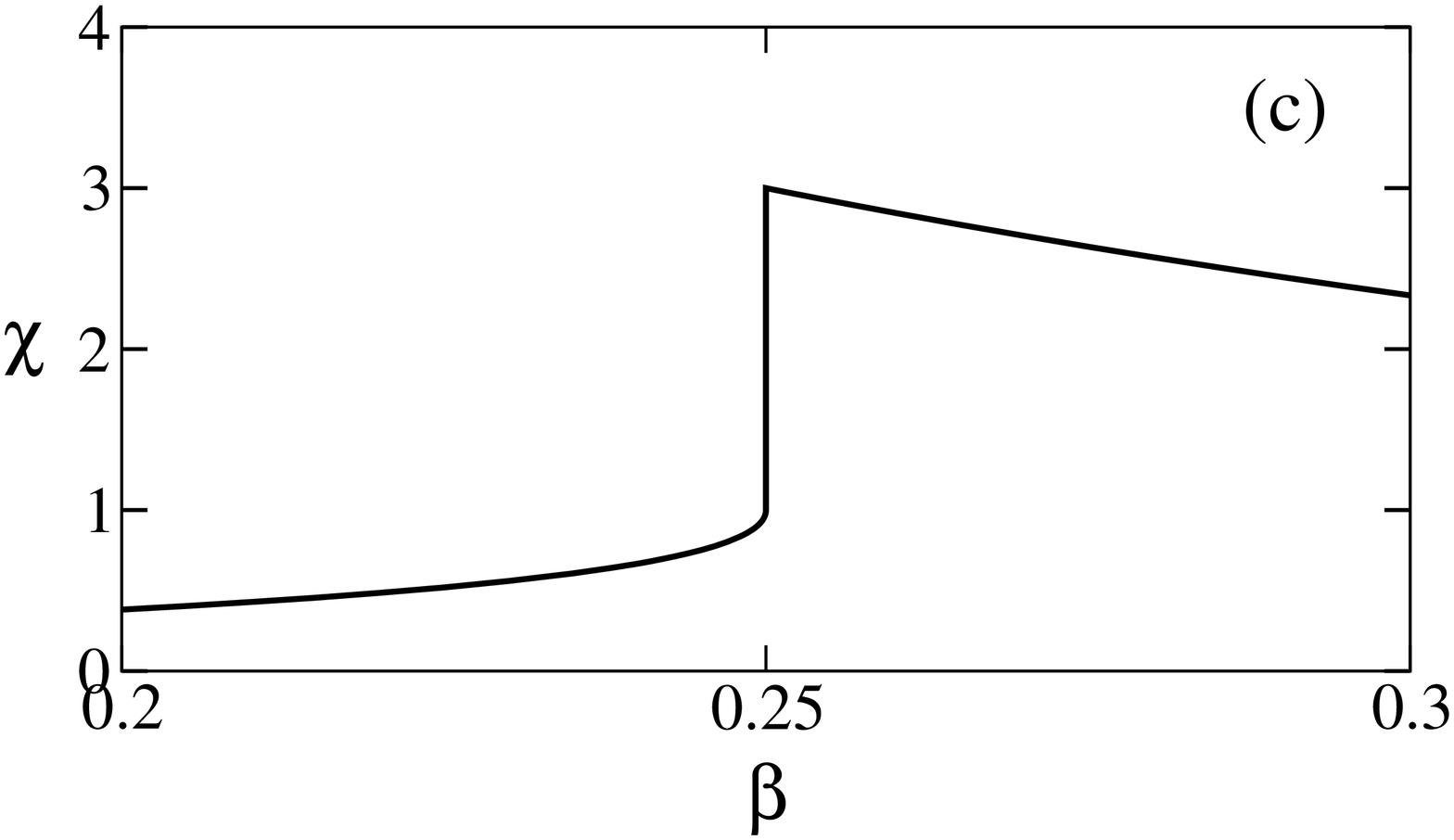}}
%\scalebox{0.28}{\includegraphics[angle=270]{degree-intervertex_fig2prime-prime.ps}}
%%%%%%%%%%%\scalebox{0.28}{\includegraphics[angle=270]{degree-intervertex_fig2a.ps}}
%%%%%%%%%%%\scalebox{0.28}{\includegraphics[angle=270]{degree-intervertex_fig2b.ps}}
\end{center}
%%\epsffile{degree-intervertex_fig2prime-prime.ps}
\caption{
Main thermodynamic quantities versus $\beta=1/T$ near the phase transition 
($\beta_c=1/4$):  
(a) the (relative) magnetization, (b) the specific heat, and (c) the magnetic susceptibility. 
}
\label{f3}
\end{figure}

%%The critical solution ($\beta_c=1/4$) has the property: $w_c(n=0)=0$. 
Near $T_c$, 
%%the phase transition point, 
the solution of Eq.~(\ref{e11}) is close to the critical one. In this range, the asymptotics of the solution at small $n$ satisfies the relation: 
\begin{eqnarray}
&&
-\frac{1}{\sqrt{4\beta-1}}\arctan \frac{[2\beta w(n)/n]-1}{\sqrt{4\beta-1}} - 
\nonumber
\\[5pt]
&&
\ln\sqrt{n^2-w(n)n+\beta w^2(n)} =\text{const}
\, 
%%,
\label{e11b} 
\end{eqnarray} 
($\beta>1/4$). This asymptotics and the solution 
%%of Eq.~(\ref{e11}) 
at large $n$ can be sewed together (see details 
%%in Ref.~\cite{dms01} and 
in the full version of the present work). 
For obtaining the dependence of the full
(relative) magnetization on $\beta$ near the critical $\beta_c=1/4$, we use the following procedure. 
(i) We substitute the boundary condition $w(n=\beta M)=M$ into relation (\ref{e11b}). 
After expansion of the arctangent, we obtain the left-hand side of the relation below:   
\begin{eqnarray*}
&&
\!\!\!\!\!\!\!-\frac{\pi/2}{\sqrt{4\beta-1}} + 1 - \ln{\frac{M(\beta)}{4}} 
= 
%%\frac{1}{1-w(n)/(2n)}
%%-\ln{[n-w(n)/2]}
\nonumber
\\[5pt]
&&
\!\!\!\!\!\!\!\frac{\pi/2}{\sqrt{4\beta\!-\!1}} - \!\!\left[1\!-\!\frac{w(n)}{2n}\right]^{-1}\!\!\!\!\! 
- \ln{\!\left[n\!-\!\frac{w(n)}{2}\right]}
\to  \frac{\pi/2}{\sqrt{4\beta\!-\!1}} -\ln{c}
%%\,   
.
\end{eqnarray*} 
(ii) On the other hand, near $\beta=1/4$, in the region $\beta M \ll n \ll 1$,  the main contribution of Eq.~(\ref{e11b}) gives the right-hand side of the equality above. 
We also use the fact that the solution must approach the critical one as $\beta\to 1/4$. So, we obtain 
the full magnetization near $\beta_c=1/4$: 
%%$M(\beta)$ in the form: 
%%
%%($\beta>1/4$) and the solution 
%%%%of Eq.~(\ref{e11}) 
%%at large $n$ can be sewed together (see details 
%%%%in Ref.~\cite{dms01} and 
%%in the full version of the present work). 
%%%%As one of the results of these procedure, 
%%Eq. (\ref{e11b}), applied respectively to $n=n(z=0)$ and to $n$
%%small but arbitrary, yields after expansion of $\arctan$ to first order 
%%($\beta \approx 1/4$)
%%%%
%%$$
%%-\frac{\pi}{2\sqrt{4\beta-1}}+1-\ln{\frac{M}{4}
%%  }=\frac{\pi}{2\sqrt{4\beta-1}}-\frac{1}{f(n)}-\ln{[nf(n)]},
%%$$
%%
%%where $\beta$ and $w(n)$ have been replaced by their critical value in 
%%the subdominant terms. Thus, we find the dependence of the full
%%(relative) magnetization on $\beta$ near the critical $\beta_c=1/4$: 
%%
\begin{equation} 
M(\beta) \cong 4ce \exp\left( -\,\frac{\pi}{2\sqrt{\beta-1/4}} \right)
\, ,     
\label{e12}
\end{equation}
where $4ce=16.90\ldots$, 
$e$ is Euler's number. 
%%is a numerical constant. 
Note that this BKT behavior is a direct result of the specific singular form of Eq.~(\ref{e11}) at small $n$ and $w$. 
%%[see Eq.~(\ref{e11a})]. 
The behaviors of the magnetization and other main thermodynamic quantities near the phase transition are shown in Fig.~\ref{f3}. 

%%%%%%%%%%%%%%%%%%%%%%%%%%%%%%%%%%%%%%%%%%%%%%%%%%%%%%%%%%%%%%%%%%%%
%%%%%%%%%%%%%%%%%%%%%%%%%%%%%%%%%%%%%%%%%%%%%%%%%%%%%%%%%%%%%%%%%%%%

%%%%%%%%%%%%%%%%%%%%%%%%%%%%%%%%%%%%%%%%%%%%%%%%%%%%%%%%%%%%%%%%%%%%
%%%%%%%%%%%%%%%%%%%%%%%%%%%%%%%%%%%%%%%%%%%%%%%%%%%%%%%%%%%%%%%%%%%% 

By using Eq.~(\ref{e11}), one can also find 
%%the dependence of 
the coefficient 
%%$A$ 
of the term $x^{2\beta}$ in relation (\ref{e7}). 
%%In the range of $x$ where $1-m(x)\ll 1$, near the phase transition, 
Near $T_c$, at small $x$, 
\begin{equation} 
m(x) \cong 1 - 2\, e^{\beta[(2\pi/\sqrt{\beta-1/4}) - 13.06]}\, x^{2\beta}
\, .     
\label{e13}
\end{equation} 
Here $H=0$. 
That is, as the temperature approaches $T_c$, $m(x)$ decreases with $x$ more and more rapidly. 

{\em Specific heat and susceptibility.}---Substituting result (\ref{e12}) into formula (\ref{e6}) for the free energy 
%%($H=0$) 
readily gives the specific heat, $tC(T) = - T\partial^2 F/\partial T^2$. 
$C(T>T_c)=0$, as is usual for 
%%situations where 
mean-field theories. 
%%is valid. 
If $T<T_c$, 
\begin{equation} 
C(T) = \frac{(\pi ce)^2}{8\,(\beta-1/4)^3} \exp\left( -\,\frac{\pi}{\sqrt{\beta-1/4}} \right)
\, ,      
\label{e15}
\end{equation} 
where $(\pi ce)^2/8=22.01\ldots$. 

Similarly, one can consider the case of a non-zero homogeneous magnetic field. 
Here we present the resulting expressions for the magnetic susceptibility: 
\begin{eqnarray} 
%%&&
\chi(\beta>1/4)& = &\beta^{-1} - 1 
\, , 
\nonumber
\\[5pt]
%%&&
\chi(\beta<1/4) &= &(1-\sqrt{1-4\beta})/(1+\sqrt{1-4\beta})
%%\frac{1-\sqrt{1-4\beta}}{1+\sqrt{1-4\beta}}
\, .     
\label{e16}
\end{eqnarray} 
There is a finite jump of the susceptibility at the phase transition point: 
$\chi[\beta\!=\!(1/4)^-]\!=\!1$ and $\chi[\beta\!=\!(1/4)^+]\!=\!3$ [see Fig.~\ref{f3}(c)]. 
%Note that with the magnetic field applied, 
%$m(x=0) = 1$ 
%%$m(x=0)$ becomes equal to $1$ 
%even at $T \geq T_c$: at small $x$ and $H$, 
%% 
%\begin{equation} 
%m(x) = 1 - B(\beta) H^{-4\beta/(1-\sqrt{1-4\beta})} x^{2\beta}
%\, ,      
%\label{e16a}
%\end{equation} 
%% 
%where 
%%the coefficient $B$ depends on $\beta$. At 
%at $T=T_c$, $B \approx 2.2$.

{\em Response to a local magnetic field.}---In networks, instead of correlations in space, one has to consider other options. 
In our case, to characterize the decrease of correlations between interacting spins, we use the distribution of linear responses to local magnetic fields. 
We apply a small field to the neighborhood of some point $y$: 
\begin{equation}
H(x,y) = h [\theta(x-(y-\Delta/2)) - \theta(x-(y+\Delta/2))] 
\, 
%%, 
\label{e17}
\end{equation} 
[$\theta(x)$ is the theta-function, $h$ and $\Delta$ are small] and find the change of the full 
%%(relative) 
magnetization which is induced by this field: 
$\mu(y) = \int_0^1 dx\,[m(x,y)-m(x)]$. Knowing $\mu(y)$ readily gives the distribution $P(\mu)$ of the response.  

Calculations are especially simple at $T>T_c$. One can find $m(x,y)$ by iterating Eq.~(\ref{e3}), which 
gives  
%%allows us to obtain $\mu(y)=M(y)$: 
%%
\begin{equation}
\mu(y) = \beta h\Delta
\frac{2}{1+\sqrt{1-4\beta}}\, y^{-(1-\sqrt{1-4\beta})/2} 
\, .
%%, 
\label{e18}
\end{equation} 
Note the power-law divergence of the linear response as $y\to0$. Relation (\ref{e18}) results in the power-law response distribution\vspace{-1pt}: 
\begin{equation}
P(\mu) \propto \mu^{-[1 + 2/(1-\sqrt{1-4\beta})]}
\, .  
\label{e19}
\end{equation} 
It is important that in contrast to ``normal'' continuous phase transitions, $P(\mu)$ is a power-law function in the entire phase and not only at $T_c$. 
As is natural, in the other phase, this distribution is a rapidly decreasing function. 
At the phase transition point, $P(\mu,\beta=1/4) \propto \mu^{-3}$.

{\em Discussion and conclusions.}---Several points must be emphasised. 

%%%%%%%!!!!!\\ 
%%\indent
%%%%%%%!!!!!$\phantom{|}\ \, $
(i) The spins on the oldest vertices 
%%usually 
are 
%%practically always 
oriented in most of situations: $m(x\!=\!0)=1$ 
%%in the entire low-tem\-pe\-rature phase and also 
even above $T_c$, if any non-zero (positive) magnetic field is applied 
at least to one spin of the system. 
%%We used the coupling 
%%Note that it is the region of relatively old vertices that is important ...
%%Note that for our conclusions only 
%%it is only the small x form of the 1/x kernel that is important

(ii) We considered networks with a specific annealing.  
The problem of a quenched disorder is more complex. However, there is a quenched situation, to which 
%%, we believe, 
our 
results are applicable. Let each new vertex have a large number $N$---greater than the final size of the network or of this order---of new connections to randomly chosen vertices. Let each of this edges bear the Ising coupling equal to $1/N$. Then we arrive at the situation similar to that is shown in Fig.~\ref{f1}. 

%%%%%%%!!!!!\\ 
%%\indent
%%%%%%%!!!!!$\phantom{|}\ \, $
(iii) The phase transition found in this paper, as well as the structural transition considered in Refs.~\cite{chk01,dms01,kkkr02,l02,bb03,d03,cb03,br05,kd04}, seriously differs from the usual BKT transition. 
%%We indicate two issues. 
%%(a) In our case, critical fluctuations are absent. In the normal BKT transition, critical fluctuations are extremely strong. (b) 
%%%%%%%%%%%%%%%%!!!!!In our case, the analogue of the power-law correlations takes place in the high-temperature phase. In contrast, in the normal BKT transition, the power-law decrease of correlations is in the low-temperature phase. 
In our case, the analogue of the power-law correlations takes place in the normal phase. In contrast, in the traditional BKT transition, the power-law decay of correlations is in the phase with a non-zero order parameter.  

%%%%%%%!!!!!\\ 
%%\indent
%%%%%%%!!!!!$\phantom{|}\ \, $
(iv) 
We stress that the more traditional-looking transitions of Refs. \cite{ahs02,dgm02,lvv02,g03,cah02,pv01} are realized in equilibrium networks where all vertices are statistically equivalent. In contrast, the networks, where we observed the transition with the BKT singularity, are specifically inhomogeneous. 
For our general conclusions, the specific $1/j$ form of the inhomogeneity of the interaction in the Hamiltonian (\ref{e1}) is important only in the region of relatively small $j$. Deviations from this form at larger $j$ do not change the critical behavior.   
%%The inhomogenety of the Ising interaction in the Hamiltonian (\ref{e1}) is present in the form $1/j$. Note however that for our general conclusions ...
%%We use the inhomogenety $1/j$ of the Ising interaction 
%%In the Hamiltonian (\ref{e1}), the inhomogeneity of the 
%%We used the coupling 
%%Note that it is the region of relatively old vertices that is important ...
%%Note that for our conclusions only 
%%%%%%%%%%%%%%%!!!!!We studied a growing network but the problem has been reduced to the Ising model on a compact system with strong inhomogeneity. We believe that our results are applicable to other systems of this kind. 
We studied a growing network but the problem has been reduced to the Ising model on a compact system with strong long-range inhomogeneity. We believe that our results are applicable to other systems with inhomogeneity of this kind. 
Furthermore, the Ising model is only a simple example of cooperative models were the observed transition should be present.  

%%%%%%%!!!!!\\ 
%%\indent
%%%%%%%!!!!!$\phantom{|}\ \, $
In conclusion, 
we have solved 
%%considered 
the ferromagnetic Ising model on a highly inhomogeneous growing net. 
%%network. 
In this system we have found an infinite order phase transition with the BKT singularity. This transition separates a phase, where the distribution of the response to a local field  is power-law, and the phase, where this distribution is 
rapidly decreasing. 
%%a rapidly decreasing function. 
We suggest that this 
%%phase 
transition also occurs in other cooperative models on compact substrates with strong long-range inhomogeneity. 
This work was partially supported by projects 
POCTI. 
%%%%: FAT/46241, 
%%/2002, 
%%%%MAT/46176, 
%%/2002, 
%%%%FIS/61665, 
%%/2004, 
%%%%and BIA-BCM/62662. 
%%/2004. 
%%SND and JFFM 
%%acknowledge .... 
%%the NATO program OUTREACH for support. JGO
%%acknowledges financial support ..... 
%%of FCT (Portugal), grant No. SFRH/BD/14168/2003. 
SND thanks A.~Samukhin for useful discussions and 
Service de Physique Th\'eorique, 
%%%%de Saclay CEA/DSM/SPhT, Unit\'e de recherche associ\'ee au 
CEA-Saclay for 
%%their 
hospitality\vspace{-0pt}. 
%%Authors are grateful to 
%%%%D.~Bernard and 
%%A.N.~Samukhin 
%%for useful discussions. 
%%SD thanks Service de Physique Th\'eorique, 
%%%%de Saclay CEA/DSM/SPhT, Unit\'e de recherche associ\'ee au 
%%CEA-Saclay for their hospitality.
%%\vspace{-8pt}. 
%%\\
%%%%\end{acknowledgments}
%%$\phantom{.}$
%%
%%
%%\vspace{-20pt}$\phantom{.}$
%%%%%%%%%%%%%%%%%%%%%%%%%%%%%%%%!!!!!\vspace{-36pt}


\begin{thebibliography}{10}
\bibitem{remark1} 
The complex architectures of the equilibrium networks may change the values of critical exponents and increase the order of the phase transition. 
However, they do not change dramatically the type of critical singularity.
\end{thebibliography}
\end{document}